\documentclass[twocolumn]{aastex61}
\usepackage{xspace}
\newcommand{\n}{\noindent}
\newcommand{\ul}{ULTRASAT\xspace}
\newcommand{\vs}{v_{s8.5}}
\newcommand{\Ni}{$^{56}$Ni\xspace}

\begin{document}

\title{Exploring the efficacy and limitations of shock-cooling models: new analysis of Type II supernovae observed by the Kepler mission}
\author{Adam Rubin and Avishay Gal-Yam}
\affil{Department of Particle Physics and Astrophysics, Weizmann Institute of Science, 234 Herzl St., Rehovot, Israel}
\email{adam.rubin@weizmann.ac.il}

\begin{abstract}

	Modern transient surveys have begun discovering and following supernovae (SNe) shortly after first light---providing systematic measurements of the rise of Type II SNe. We explore how analytic models of early shock-cooling emission from core-collapse SNe can constrain the progenitor's radius, explosion velocity, and local host extinction. We simulate synthetic photometry in several realistic observing scenarios and, assuming the models describe the typical explosions well, find that ultraviolet observations can constrain the progenitor's radius to a statistical uncertainty of $\pm10-15\%$, with a systematic uncertainty of $\pm 20\%$. With these observations the local host extinction ($A_V$) can be constrained to a factor of two and the shock velocity to $\pm 5\%$ with a systematic uncertainty of $\pm 10\%$. We also re-analyze the SN light curves presented in \citet{garnavich_shock_2016} and find that KSN 2011a can be fit by a blue supergiant model with a progenitor radius of $R_s < 7.7 + 8.8 {\rm (stat)} + 1.9     {\rm (sys)}R_\odot$, while KSN 2011d can be fit with a red supergiant model with a progenitor radius of $R_s = 111_{-21 {\rm (stat)}}^{+89 {\rm (stat)}} \phantom{}_{-1{\rm (sys)}}^{+49 {\rm (sys)}}$. Our results do not agree with those of \citet{garnavich_shock_2016}. Moreover, we re-evaluate their claims and find that there is no statistically significant evidence for a shock breakout flare in the light curve of KSN 2011d.

\end{abstract}

\keywords{}

\section{Introduction}

\label{sec:intro}

Modern surveys such as the Palomar Transient Factory \citep[PTF, iPTF;][]{law_palomar_2009,kulkarni_intermediate_2013}, the Panoramic Survey Telescope \& Rapid Response System \citep[PanSTARRS;][]{kaiser_pan-starrs:_2002}, the All-Sky Automated Survey for SuperNovae \citep[ASASSN;][]{shappee_man_2014}, the Subaru HSC Survey Optimized for Optical Transients \citep[SHOOT;][]{tanaka_rapidly_2016}, and the High Cadence Transient Survey \citep[HITS;][]{forster_high_2016} have successfully been discovering and following SNe close to their date of first light. In addition to a handful of individual objects \citep{pastorello_sn_2006,quimby_sn_2007,gezari_probing_2008,schawinski_supernova_2008,gal-yam_real-time_2011,arcavi_sn_2011,ergon_optical_2014,valenti_first_2014,bose_sn_2015,gall_comparative_2015,arcavi_constraints_2017,tartaglia_progenitor_2017}, samples with good coverage during the rise of Type II SNe have only recently been published \citep{rubin_type_2016}. \citet[G16]{garnavich_shock_2016} published two SNe discovered in the Kepler mission data. These are extremely well sampled SN II LCs and we address them in this paper. 

In parallel, theorists have developed models to describe the expected early-time emission from core-collapse SNe. While hydrodynamic models provides detailed calculation of the explosion, they are computationally expensive. Analytic models are more appropriate for large parameter space searches such as the ones performed in this work. For recent reviews of SN modeling see \cite{hillebrandt_physics_2011} and also the introduction of \cite{morozova_numerical_2016}.

\citet{waxman_grb_2007} and \citet[NS10]{nakar_early_2010} derived similar models describing the post-shock envelope emission from massive envelopes. \citet[RW11]{rabinak_early_2011} extended the theory to non-constant opacity, and improved the calculation of the color temperature by taking into account bound-free absorption which was previously neglected. 

\citet[SW17]{sapir_uv/optical_2017} re-derived the analytical results for constant opacity, and extended the theory to later times. \citet[S16]{shussman_type_2016-1} explored calibrating analytical models against numerically simulated explosions and progenitors, and also extended the theory to later times when the photosphere has penetrated more deeply into the ejecta. Both S16 and the extended theory in SW17 depend more strongly on the assumptions on the internal structure of the progenitor than the unextended theories. Therefore we limit ourselves to the unextended analytical theories. These theories depends explicitly (or implicitly) on the following assumptions:

\begin{enumerate}
	\item The ejecta has expanded sufficiently such that it is no longer planar and must be considered in the spherical geometry. NS10 and S16 do not assume this and give solutions for the LC including the planar phase.
	\item The emission is from a very thin shell which was initially near the edge of the star and the photosphere has penetrated only a small fraction of the ejected mass. This is assumed by the unextended theories NS10/RW11/SW17, but not by the extensions in S16/SW17.
	\item The temperature is above $0.7$ eV and recombination effects are not important. This is assumed by all of the theories.
\end{enumerate}

While NS10 and RW11/SW17/S16 roughly agree on the bolometric luminosity in the spherical phase, RW11, SW17, and S16 included bound-free (the dominant) absorption in the calculation of the color temperature. This can have a dramatic effect on the estimation of the progenitor's radius. Due to the above considerations we use the unextended models presented in SW17. For a more thorough discussion see SW17 Section 1. 

Several recent works \citep{gonzalez-gaitan_rise-time_2015,gall_comparative_2015,garnavich_shock_2016} compared observations to such models, but applied them at times when the models are no longer valid ($T<0.7$ eV or the photosphere has penetrated deep into the ejecta). In those works, the light curve parameters were estimated by comparing model time to peak to the rise-time of the light curves. Models with $10-15$ M$_\odot$ ejecta are valid only until $\sim5-7$ days after explosion, while they peak at $\sim12-14$ days depending on the parameters. \citet{rubin_type_2016} showed that including data beyond the model's validity range leads to incorrect assessment of the uncertainties and potentially to the acceptance of models which should be rejected.

Here we explore the potential of shock-cooling models to constrain the progenitor's radius, the explosion velocity, and the local host extinction under simulated observing programs with various facilities. We also revisit the G16 Kepler SNe and re-analyze the data while taking into account the limitations of the models and their uncertainties.

\section{The model}
\label{sec:model}
In this work we use the recent derivation for constant opacity of SW17. SW17 extended the previous models for low-mass envelopes, where the photosphere penetrates the envelope before the temperature has dropped below 0.7 eV. They found an approximation for extending the light curve (LC) after equation 2 (Section \ref{sec:intro}) no longer holds, which depends on the density structure of the star. In this work we consider stars with massive hydrogen envelopes, therefore we use the unextended model.

The two equations which we use are for the photospheric temperature and bolometric luminosity. They are given in SW17 (their equation 4) and are reproduced here:

\begin{eqnarray}
\label{eq:Tph}
T_{ph} & = & 1.61\;[1.69]\left(\frac{\vs^2 t_{\rm d}^2}{f_\rho M_0\kappa_{0.34}}\right)^{\epsilon_1} \frac{R_{13}^{1/4}}{\kappa^{1/4}_{0.34}}t_{\rm d}^{-1/2}\,{\rm eV} \\
\label{eq:L}
L & = &  2.0\;[2.1]\times10^{42} \left(\frac{\vs t_{\rm d}^2}{f_\rho M_0\kappa_{0.34}}\right)^{-\epsilon_2} \frac{\vs^2 R_{13}}{\kappa_{0.34}}\,{\rm \frac{erg}{s}}
\end{eqnarray}

\n where $\kappa=0.34\kappa_{0.34}\;{\rm cm^2\; g^{-1}}$, $v_{\rm s*}=10^{8.5}\vs\;{\rm cm\; s^{-1}}$, $M=1M_0\;M_\odot$, $R=10^{13}R_{13}\;{\rm cm}$, $\epsilon_1=0.027\;[0.016]$,  and $\epsilon_2=0.086\;[0.175]$ for $n=3/2\;[3]$. $v_{s*}$ is the asymptotic shock velocity, $M_0$ is the ejected mass, and $t_d$ is the time in days. 

The model is valid for the following times:

\begin{eqnarray}\label{eq:t_limits}
   t&>&0.2\frac{R_{13}}{\vs}\max\left[0.5,\frac{R_{13}^{0.4}}{(f_\rho\kappa_{0.34}M_0)^{0.2}\vs^{0.7}} \right]\,{\rm d} \\
   t&<& 3f_\rho^{-0.1}\frac{\sqrt{\kappa_{0.34}M_0}}{\vs}\,{\rm d}
\end{eqnarray}

\n where the first limit describes the requirement for sufficient expansion (spherical phase) and the second limit describes the requirement that the photosphere has only penetrated a small fraction of the envelope's mass. Additionally, to ensure fully ionized hydrogen we require

\begin{equation}
	t<{\rm arg}(T_{ph}(t)=0.7 {\rm eV})
\end{equation}

The specific flux is given by

\begin{equation}
	f_\lambda = \frac{L_{bol}}{4\pi R_{ph}^2} \frac{T_{col}}{hc} g_{BB}\left(\frac{hc}{\lambda T_{col}}\right)
\end{equation}

\n where $R_{ph}$ is the photospheric radius, $T_{col}$ is the color temperature, and $g_{BB}$ is the dimensionless black body function given by

\begin{equation}
	g_{BB}(x) = \frac{15}{\pi^4} \frac{x^5}{e^x - 1}
\end{equation}

SW17 explored numerically several different progenitors with varying core to mantle mass ratios and found that the density normalization $f_\rho=1-3\;[0.1-0.8]$ for $n=3/2\;[3]$ respectively (assuming normal stars with core to mantle mass ratios of $0.1-1$, SW17 Figure 5). Here we take $f_\rho=1\;[0.1]$ for $n=3/2\;[3]$ respectively, appropriate for a core to mantle mass of $1$. However the emission is weakly dependent on $f_\rho$. SW17 also show that the ratio of the color temperature to the photospheric temperature is well behaved and given by $T_{col}/T_{ph}=1.1\;[1.0]\pm0.025[0.05]$ for $n=3/2\;[3]$ (SW17 Figures 11 and 13). We use these nominal values. See Section \ref{sec:systematics} and Figure \ref{fig:systematics} for the effect of these systematic uncertainties on the inferred parameters.

\section{Fitting synthetic data}

In order to estimate the efficacy of shock-cooling models we simulated a synthetic photometry campaign. We explored discovery in R-band 0.5 days after explosion with follow-up triggered one day later. We synthesized the following followup scenarios:  photometry in  \citet{bessell_ubvri_1990} BVI, UBVI, or BVI + \emph{SWIFT/UVOT} UVM2 with a one day cadence and R observation three times per night. We excluded UVW1 and UVW2 due to their known red leaks \citep{brown_absolute_2010}. These followup scenarios are realistic and are similar to the observational campaign of SN 2013fs \citep{valenti_diversity_2016,yaron_confined_2017}. We simulated SNe with various radii ($R_s = 50-1000 R_\odot$) and extinction values ($A_V = 0.1-1$).

In order to explore what may be achieved with future UV facilities, we also simulated the expected photometry from \ul \citep{sagiv_science_2014}. \ul is a proposed UV satellite observatory which will acquire high cadence (15 minute) UV photometry at $2500$\AA. We simulated photometry in the \ul filter with 1 hour cadence and pre-explosion photometry (a conservative scenario given the 15 minute cadence design). 

\begin{figure*}[ht]
	\centering
	\includegraphics[width=0.75\textwidth]{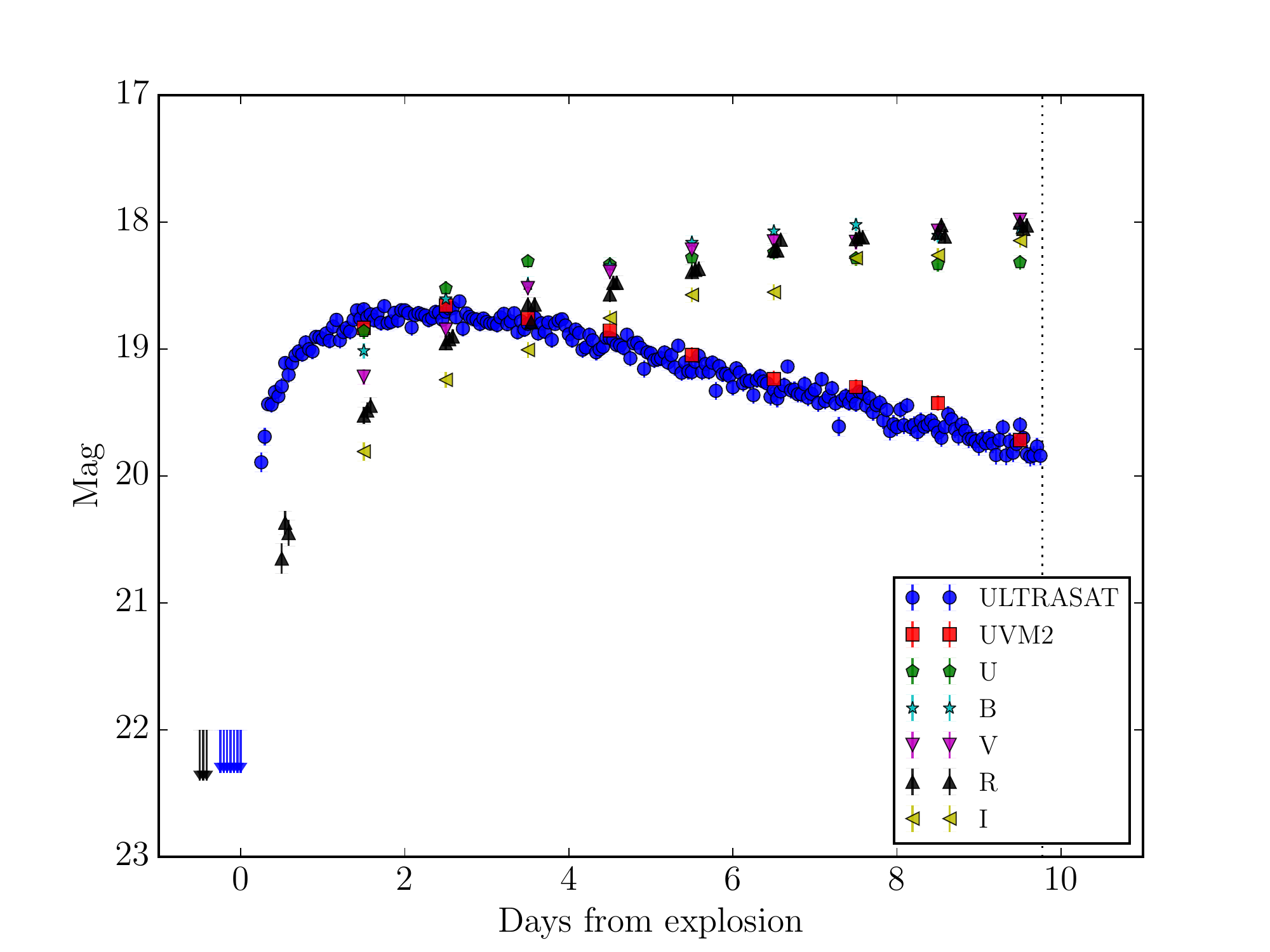}
	\caption{Example synthetic light curve generated from SW17 of a star with $R=500 R_\odot$, $A_V=0.1$ and the parameters given in Table \ref{tab:synth_parameters}. The error bars are plotted, but are smaller than the markers. The time where the model is no longer valid is marked by the dotted vertical line.}
	\label{fig:examples_lcs_r500av01v1.pdf}
\end{figure*}

\begin{figure*}[ht]
	\centering
	\includegraphics[width=0.75\textwidth]{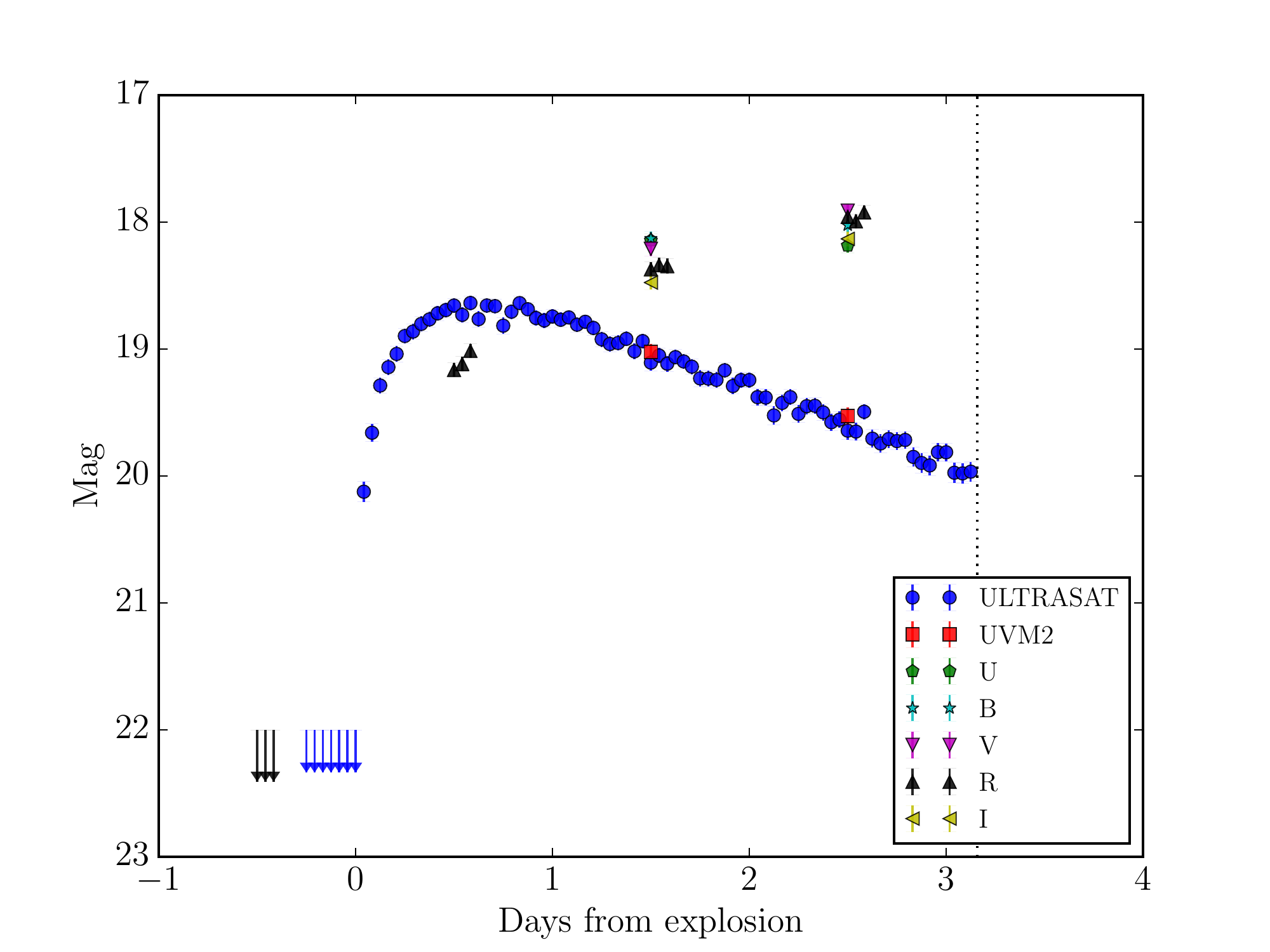}
	\caption{Example synthetic light curve generated from SW17 of a star with $R=50 R_\odot$, $A_V=0.1$ and the parameters given in Table \ref{tab:synth_parameters}. The error bars are plotted, but are smaller than the markers. The time where the model is no longer valid is marked by the dotted vertical line.}
	\label{fig:examples_lcs_r50av01v1.pdf}
\end{figure*}

We generated synthetic data from equations \ref{eq:Tph} and \ref{eq:L} using the parameters in Table \ref{tab:synth_parameters}. The extinction for each central wavelength was calculated using the \citet{cardelli_relationship_1989} prescription. Different redshifts were chosen for the models such that they gave roughly the same observed peak R magnitude ($m_{peak}\sim18$). All magnitudes reported here are in the AB system unless stated otherwise. The distance moduli were calculated using \citet{planck_collaboration_planck_2015} cosmology with $H_0=67.74$, $\Omega_m=0.31$, $\Omega_\Lambda=0.69$. Two examples of the synthetic light curves are shown in Figures \ref{fig:examples_lcs_r500av01v1.pdf} and \ref{fig:examples_lcs_r50av01v1.pdf}.

\begin{deluxetable}{lc}
	\tabletypesize{\scriptsize}
	\tablecaption{Parameters used in the synthetic data \label{tab:synth_parameters}}
	\tablewidth{0pt}
	\tablehead{\colhead{} & \colhead{}}
	\startdata
	$n$      &  3/2   \\
	$\kappa$      &  0.34   \\
	$Rs/R_\odot$  &  50-1000\\
	$M/M_\odot$  &  10\\
	$f_\rho$      &  1.0    \\
	$T_{col}/T_{ph}$ & 1.1        \\
	$\vs$         &  1.0    \\
	$A_V$         &  0.05-1 \\
	$R_V$         &  3.1    \\
	\enddata
\end{deluxetable}

\subsection{Noise model}
To get a realistic model of the noise we adopted $5\%$ uncertainties for all filters. We used a limiting magnitude of 22 in all filters, similar to the limiting magnitudes observed in the SN 2013fs campaign. The effective wavelengths and limiting magnitudes used in this work are summarized in Table \ref{tab:filt_parameters}. We converted the limiting magnitude to a flux error $\sigma_b$ and used the following equation to generate noise for the model.

\begin{equation}
	\sigma^2 = \sigma_b^2 + (0.05f)^2
\end{equation}

\n where $f$ is the model flux. We then drew the synthetic observations from a normal distribution with mean $f$ and variance $\sigma^2$.

\begin{deluxetable}{lcc}
	\tabletypesize{\scriptsize}
	\tablewidth{0pt}
	\tablecaption{Filter parameters used. \label{tab:filt_parameters}}
	\tablehead{\colhead{Filter} & \colhead{Effective wavelength} & \colhead{Limiting magnitude} \\ \colhead{} & \colhead{(\AA)} & \colhead{}}
	\startdata
	UVM2  &  2262.1     &  22.0   \\
	U     &  3605.0  &  22.0   \\
	B     &  4413.0  &  22.0   \\
	V     &  5512.1  &  22.0   \\
	R     &  6585.9  &  22.0   \\
	I     &  8059.8  &  22.0   \\
	\ul   &  2500.0  &  22.0  \\
	KEPLER   &  6416.8  &  ---  \\
	\enddata
\end{deluxetable}

\subsection{Fitting procedure}
Fitting NS10/RW11/SW17 models with a simple log-likelihood test statistic is non-trivial because for different sets of parameters the models are valid for different time durations. One possible solution, which is not satisfactory, is to limit the analysis to a specific window of time. This approach was taken in \citet{rubin_type_2016,valenti_first_2014}, and \citet{bose_sn_2015}. While this guarantees that all of the explored models are valid in the window, it does not take into account that the models must be valid over \emph{their} entire range of validity (including data points outside the chosen window). We solve this problem by considering the P-value of each fit. The process is as follows:

\begin{itemize}
	\item Choose a set of model parameters
	\item Calculate the time range of validity given the parameters
	\item Calculate the P-value using
		\begin{equation}
			P = 1 - \rm{CDF}(\chi^2,\nu)
		\end{equation}
		where CDF is the cumulative distribution function, $\chi^2$ is the value of the chi-squared statistic of the fit, and $\nu$ is the number of degrees of freedom (including only those data points which are within the time range of validity).
\end{itemize}

In this way we can ensure that the accepted fits are \emph{fully self-consistent}, meaning they fit all of the relevant data and nothing but the relevant data. We defined the critical value to be P-value$>5\%$. This means that there is less than a $5\%$ chance that the data came from a rejected model. 

We generated models on a grid of $\vs$, $Rs$, and $A_V$. For each point on the multi-dimensional grid we calculated the $\chi^2$ statistic taking into account only those data points which were at times when the model is valid. For some cases we performed a Markov-Chain Monte-Carlo (MCMC) which we confirmed gives equivalent results, but was easier to use to explore the uncertainties. Note that there is a dependence on $M_{ej}$, and $f_\rho$ through the small pre-factors in equations \ref{eq:Tph}. We discuss these in Section \ref{sec:systematics}.

To test the ability to discriminate between models with $n=3/2$ and $n=3$, we performed a Monte Carlo where we drew 200 LCs with $n=3/2$ and $R=500R_\odot$, and 200 LCs with $n=3/2$ and $R=50R_\odot$. We then fit each model once assuming $n=3/2$, and once assuming $n=3$ (the incorrect polytropic index). We collected the P-Values of the best fits and compared their distributions.

\subsection{Systematic uncertainties}
\label{sec:systematics}

SW17 models depend to some degree on underlying assumptions of the stellar structure. This appears through two parameters, $M_{ej}$ and $f_\rho$, which weakly affect the results through the exponents $\epsilon_1$ and $\epsilon_2$ in equations \ref{eq:Tph} and \ref{eq:L} and appear as the degenerate combination $M_{ej} f_\rho$. In addition, the predicted band luminosity depends on the color temperature and its relation to the photospheric temperature. This dependence is degenerate between $R$ and $\vs$. This is because the photospheric temperature depends on $R^{1/4}$ and is practically independent of $\vs$, while the bolometric luminosity goes like $\vs^2 R$. 

\subsection{Results and discussion}

\begin{figure}[ht]
	\includegraphics[width=1\columnwidth]{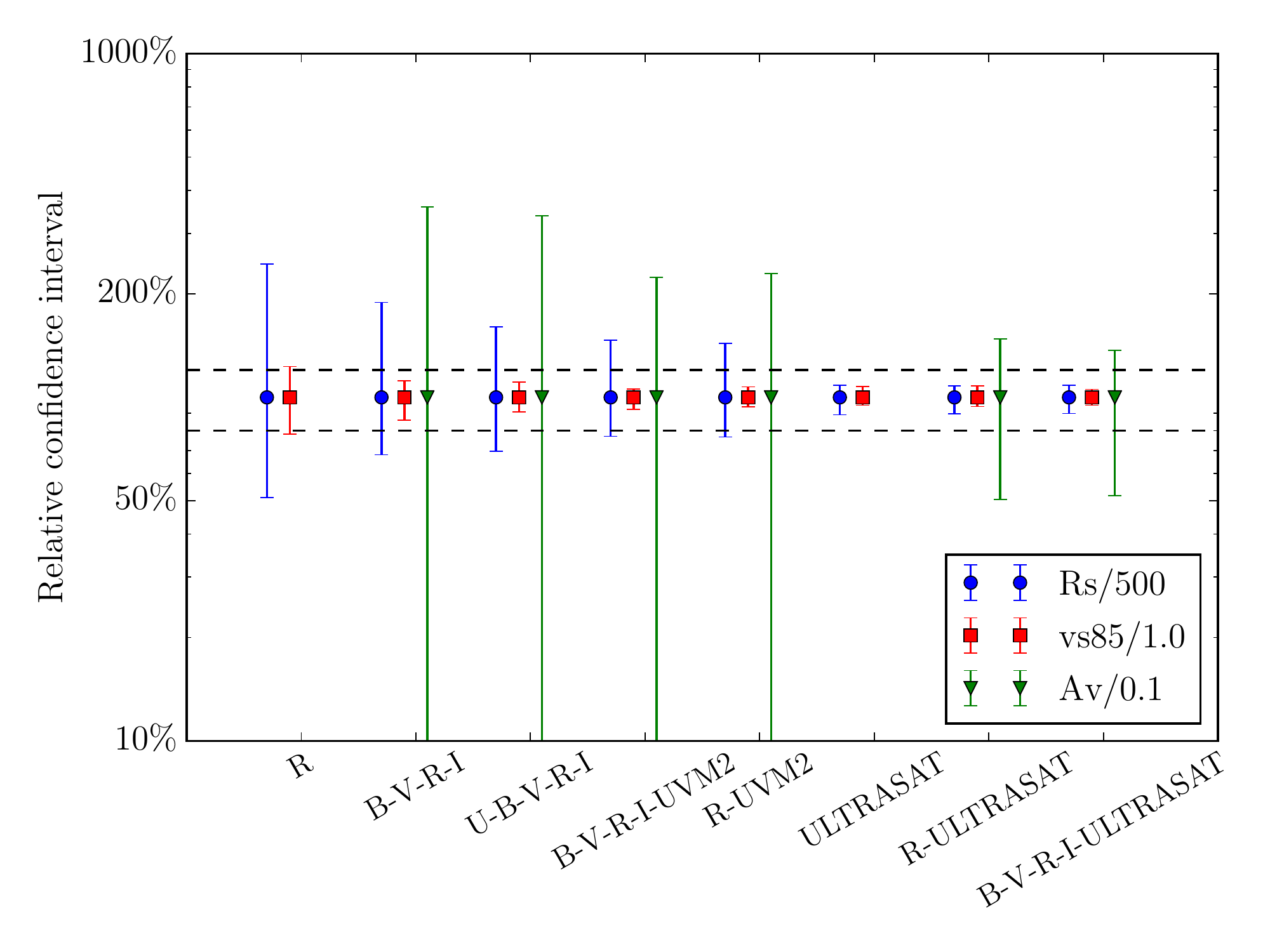}
	\caption{Relative confidence intervals for $R=500,Av=0.1,\vs=1$. $\pm20\%$ is shown in dashed lines to represent a rough estimate of the systematic uncertainty on $R$ and $\vs$.}
	\label{fig:efficacy_r500_av01_v1}
\end{figure}

\begin{figure}[ht]
	\includegraphics[width=1\columnwidth]{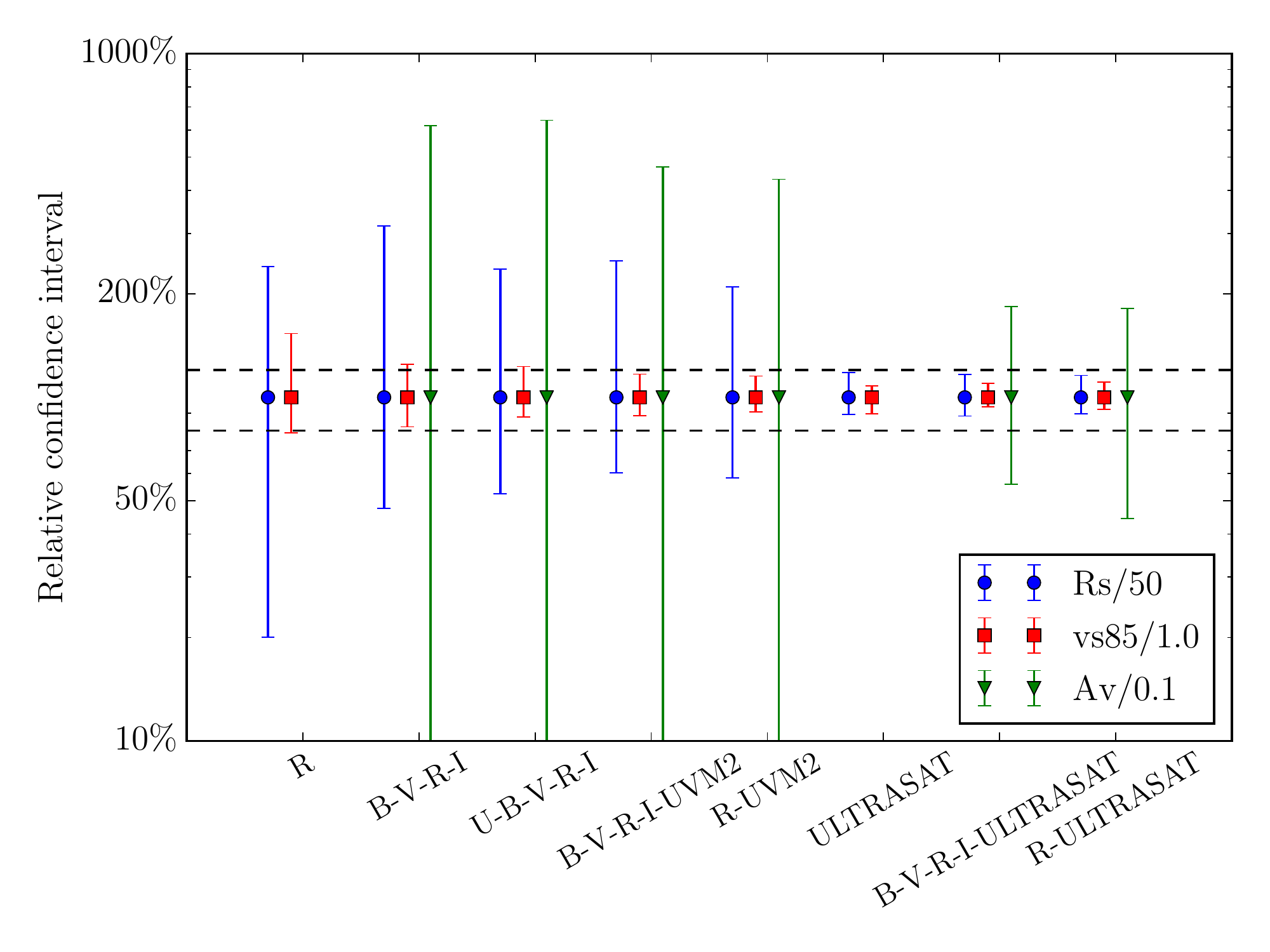}
	\caption{Relative confidence intervals for $R=50,Av=0.1,\vs=1$. $\pm20\%$ is shown in dashed lines to represent a rough estimate of the systematic uncertainty on $R$ and $\vs$.}
	\label{fig:efficacy_r50_av01_v1}
\end{figure}

Our results are summarized in Figures \ref{fig:efficacy_r500_av01_v1} and \ref{fig:efficacy_r50_av01_v1}. As was observed in \citet{rubin_type_2016}, $\vs$ ($E/M$ in their paper) is statistically well constrained. However, in single bands $\vs$ can only be considered a lower limit due to the unknown local host extinction. The addition of UV coverage reduces the uncertainties significantly. Most of the statistical power is in the UV, shown by the minor differences between R+UVM2 and BVRI+UVM2. High cadence UV reduces the uncertainties even more, to below the systematic errors for both large and small radii. It is noteworthy that for smaller radii there is a paucity of data within the valid time ranges---so a continuous, high cadence campaign is valuable.

\begin{figure}[ht]
	\includegraphics[width=1\columnwidth]{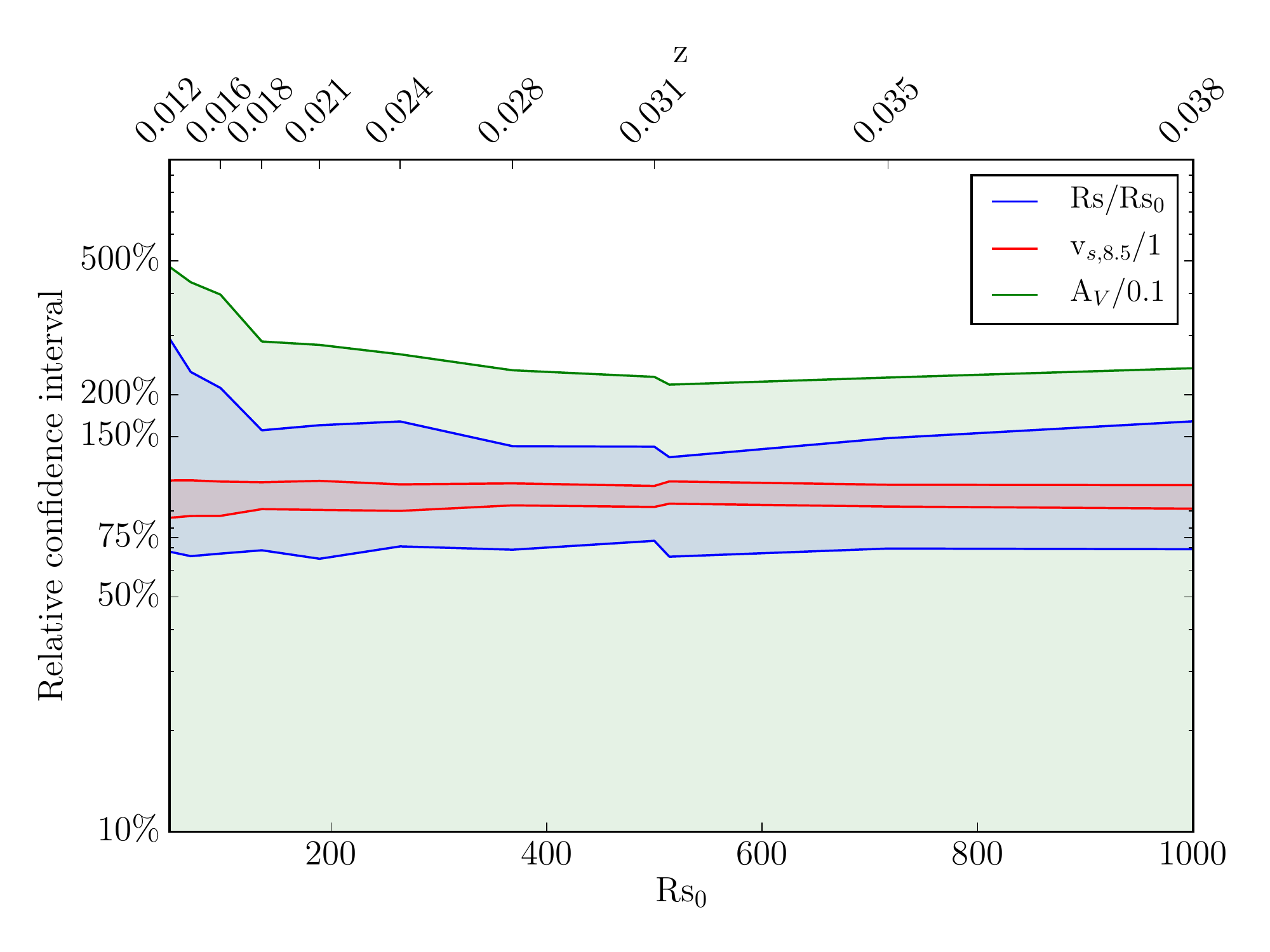}
	\caption{95\% confidence intervals for $Av=0.1,\vs=1,n=3/2$ for different radii under an observing plan of BVRI+UVM2.}
	\label{fig:ci_diff_rad}
\end{figure}

\begin{figure}[ht]
	\includegraphics[width=1\columnwidth]{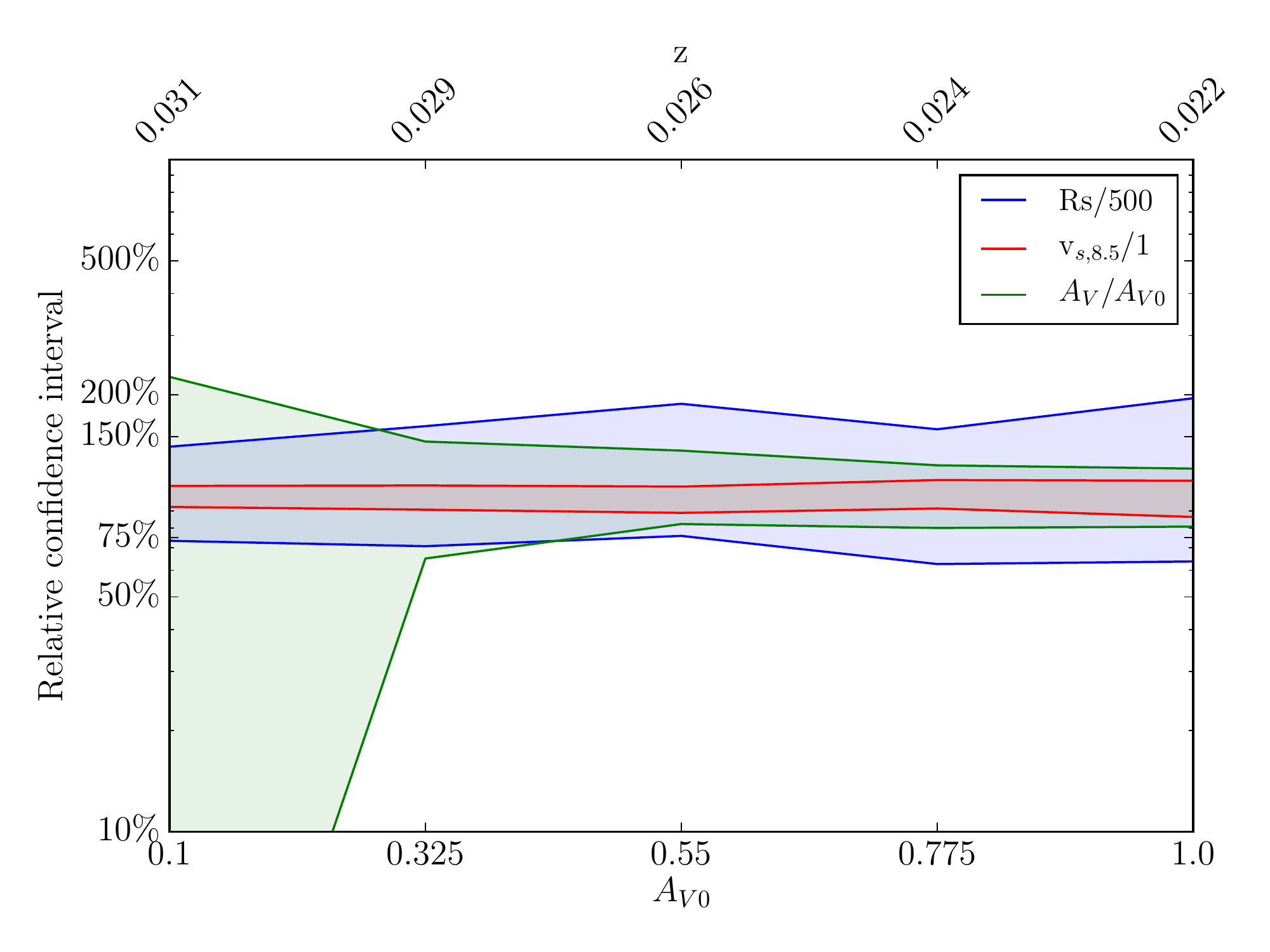}
	\caption{95\% confidence intervals for $R=500,\vs=1,n=3/2$ for different values of $A_V$ under an observing plan of BVRI+UVM2.}
	\label{fig:ci_diff_av}
\end{figure}

In Figures \ref{fig:ci_diff_rad} and \ref{fig:ci_diff_av} we show the effect of different radii and extinction $A_V$ under an observing plan of BVRI+UVM2. The relative confidence interval in radius and $\vs$ is not very sensitive to the radius or extinction for radii above $200R_\odot$. For lower radii the relative confidence interval increases, but remains less than a factor of two, indicating that the fit would still reconstruct a small radius. This sensitivity is primarily due to insufficient data in the first few days. Also, for very low values of extinction, the relative error becomes large, but the absolute upper limits are stringent at low extinction. 

\begin{figure}[ht]
	\includegraphics[width=1\columnwidth]{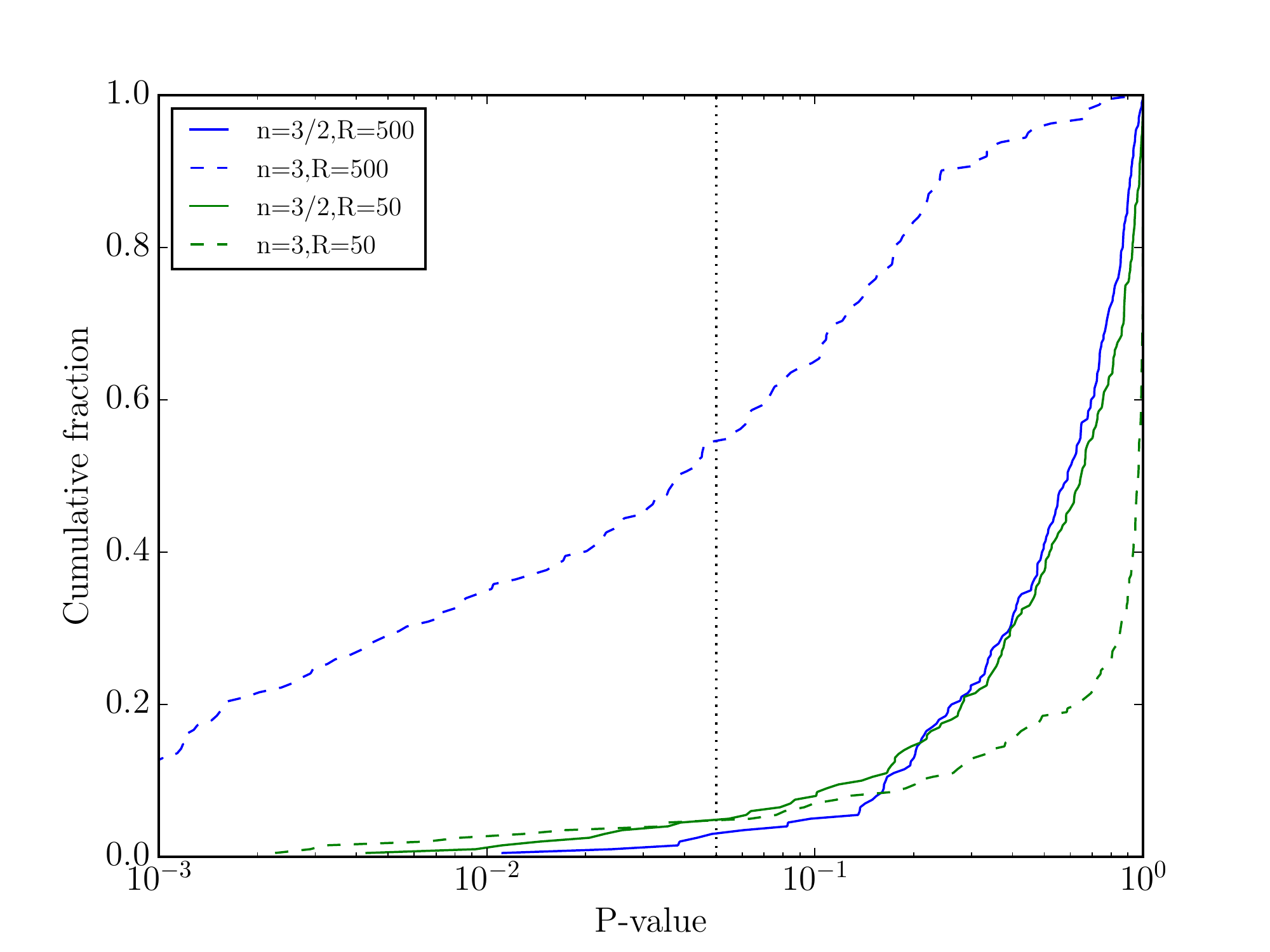}
	\caption{Monte Carlo simulations showing the P-value distribution for two $n=3/2$ models. The solid lines show the P-value distribution when fitting $n=3/2$ models assuming $n=3/2$, while dashed lines show the P-value distribution when fitting $n=3/2$ models assuming $n=3$ (the incorrect polytropic index). With the 1-day cadence follow-up campaign with BVRI+UVM2 coverage it is possible to discriminate between $n=3/2$ and $n=3$ for models with large radii $(R\sim 500 R_\odot)$, but not for models with smaller radii $(R\sim 50 R_\odot)$.}
	\label{fig:compare_n32_n3}
\end{figure}

\begin{figure}[ht]
	\includegraphics[width=1\columnwidth]{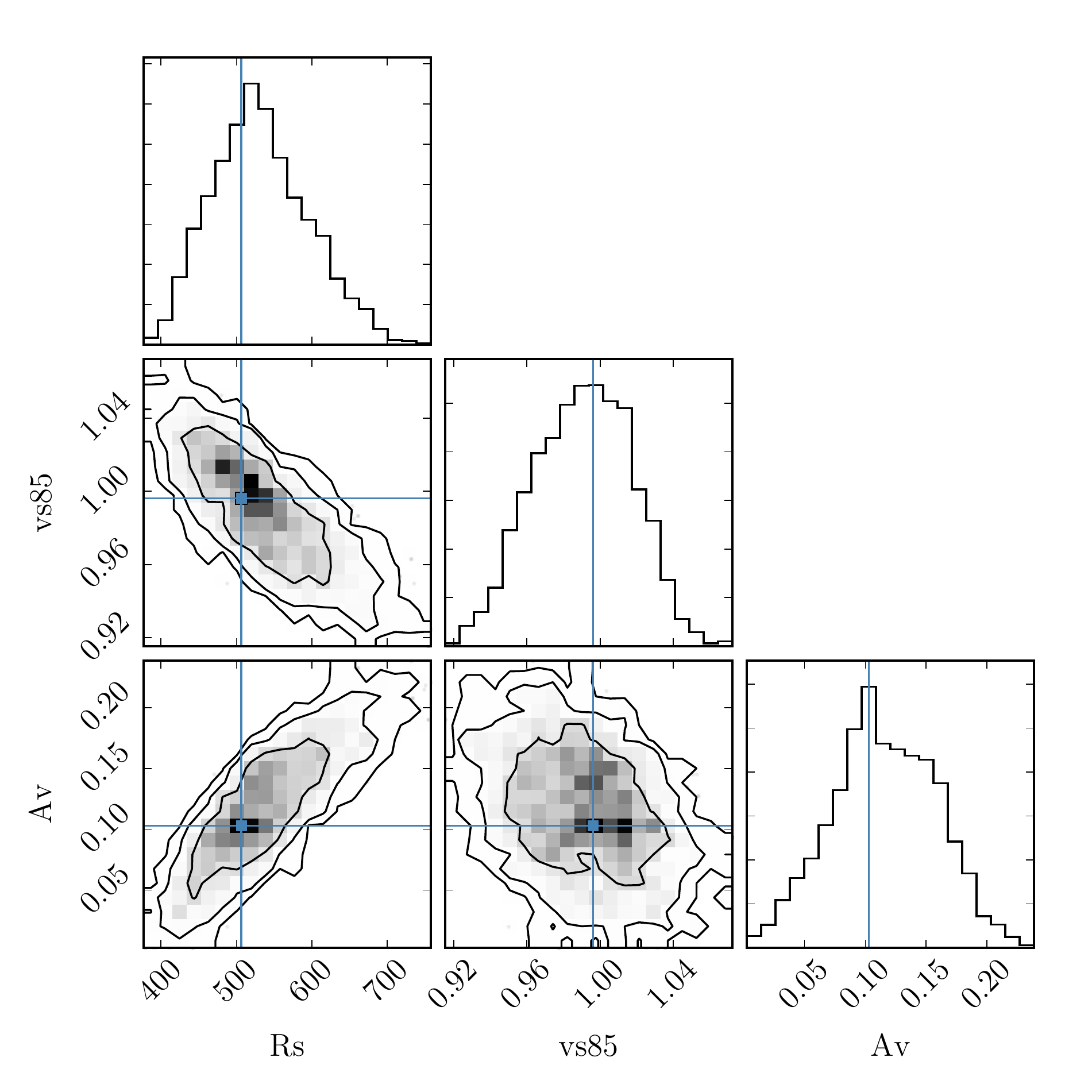}
	\caption{Distribution of the fit parameters $Rs,\vs$ and $A_V$ resulting from an MCMC fit to a SW17 model of a SN explosion with a progenitor with $R=500R_\odot,\vs=1,A_V=0.1$ and the parameters given in Table \ref{tab:synth_parameters} for an observing program of BVRI+UVM2. The correlation between $Rs$ and $\vs$, and the correlation between $Rs$ and $A_V$ are apparent, while $A_V$ and $\vs$ are uncorrelated with this observing plan. The contours represent 68\%, 95\% and 99.7\% confidence intervals. The blue marks show the result of a direct minimization.}
	\label{fig:corner_example}
\end{figure}

Figure \ref{fig:compare_n32_n3} shows the results of the Monte Carlo simulations testing how well the models can discriminate between models with $n=3/2$ and $n=3$. For large radii the models can be quite easily discriminated between, achieving a plausible fit (P-Value$>0.05$) for the incorrect polytropic index in $\sim50\%$ of the time, with much lower likelihood than the correct polytropic index. However, at lower radii the models cannot be told apart. Note, that fitting an $n=3$ model to data drawn from an $n=3/2$ model leads to larger radii (roughly by a factor of two). Figure \ref{fig:corner_example} shows an example of the 95\% confidence interval contours for $R=500R_\odot,\vs=1,A_V=0.1$ for an observing program with BVRI+UVM2. The correlation between $\vs$ and the radius, as well as the correlation between the radius and $A_V$ are noteworthy.

\begin{figure*}[ht]
	\includegraphics[width=1\textwidth]{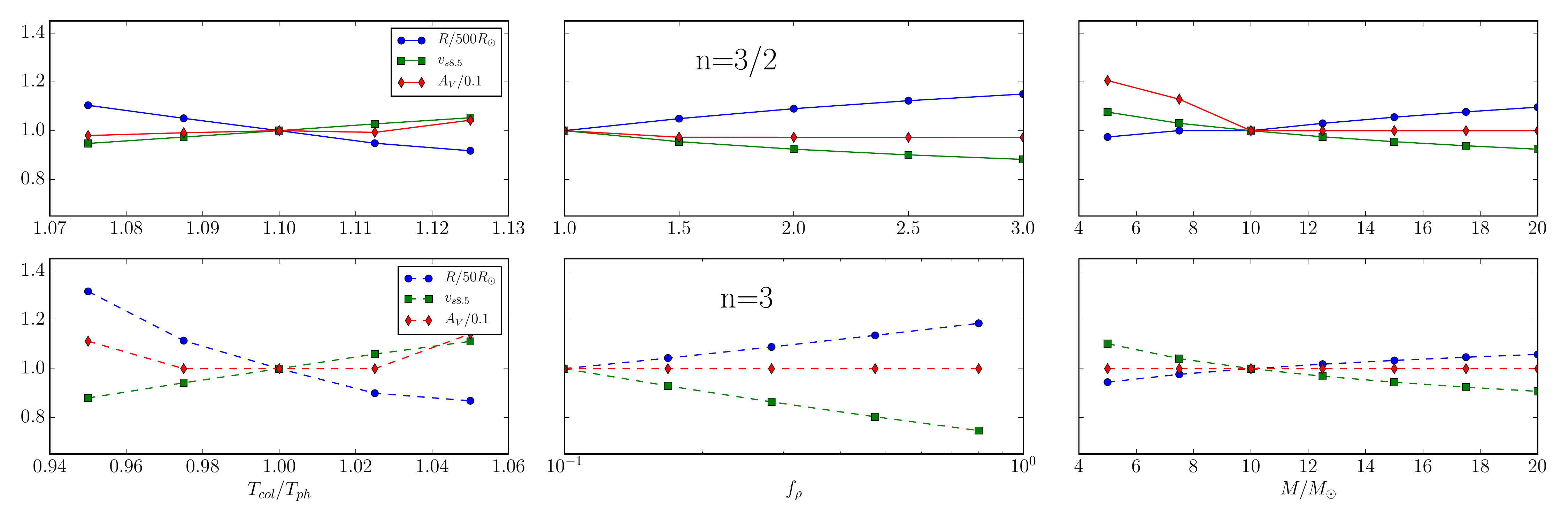}
	\caption{Top: systematic dependence of the best fit values for $R_s$, $\vs$, and $A_V$ on the unknown parameters $T_{col}/T_{ph}$, $f_\rho$, and $M_{ej}$ for $n=3/2$. Bottom: same for $n=3$ with BVRI + UVM2 coverage. For each case we used the nominal value of two parameters and varied the third.}
	\label{fig:systematics}
\end{figure*}

We evaluated the systematic uncertainties by exploring how the best fit parameters depend on the values of $T_{col}/T_{ph}$, $f_\rho$ and $M_{ej}$. The results are shown in Figure \ref{fig:systematics}. For each case we studied we fit the model to the extreme cases of the systematic uncertainties in $T_{col}/T_{ph}$, and $f_\rho$. We report the most extreme best fit values for the radius and velocity as the systematic uncertainties. The ejected mass can in principle be constrained from observations \citep{dessart_determining_2010}. Therefore we do not treat it as a systematic error, however we show in Figure \ref{fig:systematics} the effect of varying the ejected mass between $5-20$ M$_\odot$. As can be seen, the effect is weak and shifts the best fit value by roughly $\pm10\%$.

Our interim conclusions can be summarized as follows:

\begin{itemize}
	\item Single optical band coverage (e.g. R-band) cannot constrain the radius to less than a factor of two. Adding multi-band coverage (BVRI or UBVRI) reduces the uncertainty on the radius to $30\%-50\%$, and allows placing upper limits on the extinction up to a factor of four to six.
	\item High cadence UV coverage reduces the statistical uncertainty on the progenitor's radius to $\pm 10\%$. While this is currently systematics dominated, improved theories and measurements may help to further reduce this. The addition of UV coverage to measurements in the optical bands also allows for the determination of the local host extinction, which is currently challenging to determine, to within $30-100\%$.
	\item Shock-cooling models can discriminate between progenitors with $n=3/2$ and $n=3$ density profiles but this depends on the specific observing plan and cadence. Models with larger radii can be more easily discriminated between.
	\item An important caveat is the assumption of a constant reddening law, and specifically $R_V=3.1$. \citet{poznanski_improved_2009} and more recently \citet{rodriguez_photospheric_2014} and \citet{de_jaeger_hubble_2015} showed for samples of Type II-P SNe that $R_V<2$ assuming SN II are standard candles, which is a topic of debate. Also, $R_V$ for SNe may not be the same as $R_V$ for the Galaxy.
\end{itemize}

\section{Application to KSN 2011a and KSN 2011d}

KSN 2011a and KSN 2011d are two Type II-P SNe recently reported on by \citet[G16]{garnavich_shock_2016}. The parameters of both SNe are presented in Table \ref{tab:kepler_sne} (reproduced from G16). G16 analyzed their light curves with RW11 RSG models and reported their best fit parameters to be progenitor radii of $280\pm20$ R$_\odot$ and $490\pm20$ R$_\odot$ for KSN 2011a and KSN 2011d respectively, both with explosion energies of $2\pm0.3\times 10^{51}$ erg (for $M_{ej}=15$ M$_\odot$). Their analysis included light curve data until peak magnitude. G16 concluded that KSN 2011a is not consistent with the simple shock-cooling model, but requires the shock-breakout to occur from a circumstellar material. This is primarily due to the fast rise observed over a few days. KSN 2011d was well fit by the model, and G16 interpreted an excess at the very early time of the LC as a shock-breakout flare. Here we re-analyze the photometry of KSN 2011a and KSN 2011d, taking into account the limitations of the validity of the models. 

Photometry of KSN 2011a and KSN 2011d were obtained from P. M. Garnavich.\footnote{Private communication.} We binned the data into 2 hour intervals, taking the errors to be $\sigma/\sqrt{N-1}$ where $\sigma$ is the standard deviation and $N$ is the number of samples in the interval. Because only a single band is available for the Kepler SNe, we assumed no host galaxy extinction, and treat our $\vs$ as a lower limit (as did G16).

\begin{figure}[ht]
	\includegraphics[width=1\columnwidth]{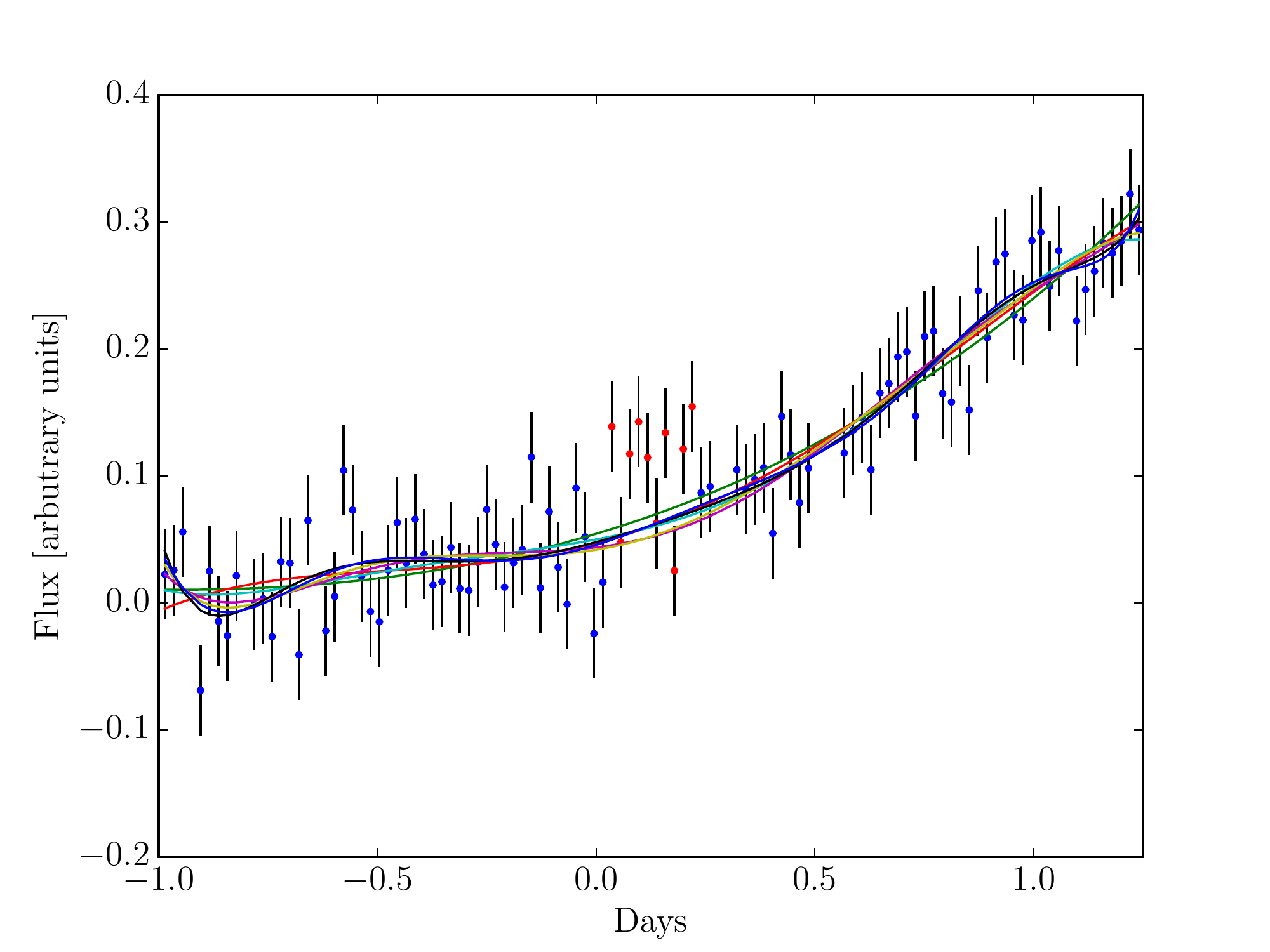}
	\caption{Light curve of KSN 2011d around the claimed shock-breakout. Red points are associated by G16 with shock breakout. The smooth curves are $3^{rd}-9^{th}$ order polynomial fits (excluding the red points). The departure of the points from the smooth curves is $3\sigma-4\sigma$ depending on the order of the fit.}
	\label{fig:sb_poly_fits}
\end{figure}

As G16 noted in their paper, there is correlated excess in the LC of KSN 2011d prior to their ``shock-breakout''. Therefore, to assess if there is a significant departure from a smooth rise, it is more reasonable to compare the light-curve to a smooth function. To describe the smooth function from which the shock-breakout may or may not depart, we fit polynomials ($3^{rd}-9^{th}$ order) to the day before and day after the ``shock-breakout'', excluding the ten points which G16 associated with it. These fits are shown in Figure \ref{fig:sb_poly_fits}. Our results are not sensitive to the polynomial degree. 

We evaluated whether or not the shock-breakout is significant in two ways. First we examined the effect of binning on the significance and shape of the departure. The native cadence of Kepler is 30 minutes, while G16 binned their data into 3.5 hr bins. This leaves seven possible phases of binning. We tested how the shape and significance of the departure differ with all possible choices of binning. Second, we measured the probability of departure of all sets of ten consecutive points in the light-curve in the data up to two days before the SN explosion. The purpose of this test is to demonstrate the ``look elsewhere'' effect, which makes a $3-4\sigma$ departure extremely likely when considering a sufficiently large amount of data, as is the case for KSN 2011d.

\begin{deluxetable*}{lccccccc}
\tabletypesize{\scriptsize}
\tablecaption{Kepler Type II-P Supernova Candidates\tablenotemark{a} \label{tab:kepler_sne}}
\tablewidth{0pt}
\tablehead{
\colhead{Name} & \colhead{Host} & \colhead{SN} & \colhead{Redshift} & \colhead{MW A$_V$} & \colhead{Peak Kp\tablenotemark{c}} & \colhead{Date of Breakout} & \colhead{Rise Time} \\ 
\colhead{} & \colhead{KIC\tablenotemark{b}} & \colhead{Type}  & \colhead{($z$)} & \colhead{(mag)}
& \colhead{(mag)} & \colhead{(BJD-2454833.0)} & \colhead{(days)}
}
\startdata
KSN 2011a & 08480662 & II-P & 0.051 & 0.194 & 19.66$\pm 0.03$ & 934.15$\pm 0.05$ & 10.5$\pm 0.4$ \\
KSN 2011d & 10649106 & II-P & 0.087 & 0.243 & 20.23$\pm 0.04$ &  1040.75$\pm 0.05$ & 13.3$\pm 0.4$ \\
\enddata
\tablenotetext{a}{Reproduced from \citet{garnavich_shock_2016}}
\tablenotetext{b}{Kepler Input Catalog \citep{brown_kepler_2011}}
\tablenotetext{c}{Not corrected for extinction}
\end{deluxetable*}

\subsection{Results and discussion}

\begin{figure*}[ht]
	\centering
	\includegraphics[width=0.75\textwidth]{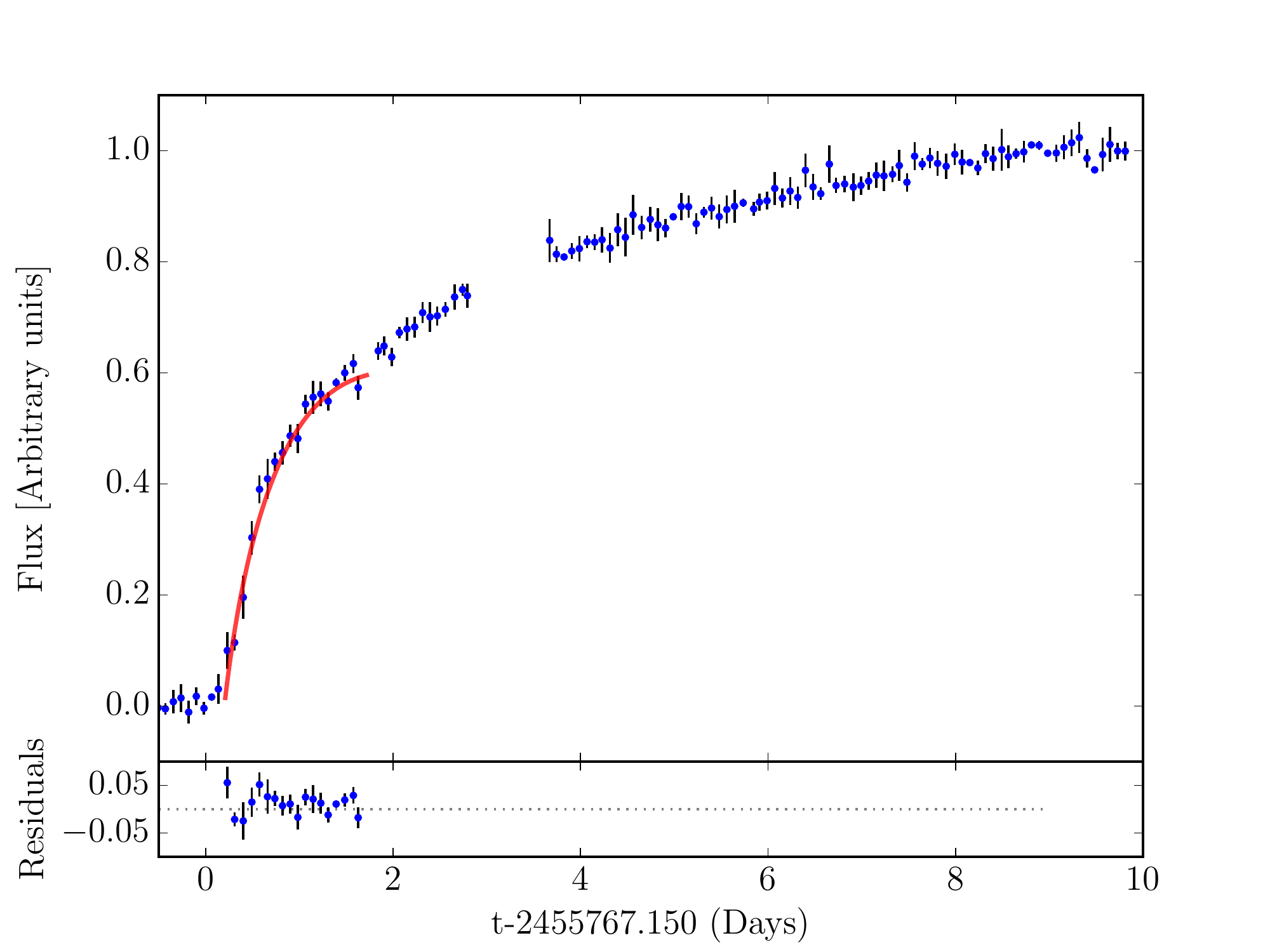}
	\caption{Top: best fit to KSN 2011a. Blue points are the 2 hour binned data. The models are valid only for the times where the best fit red line is drawn. Bottom: residuals.}
	\label{fig:ksn11a_bestfit}
\end{figure*}

Our results for the fit parameters of KSN 2011a and KSN 2011d are presented in Table \ref{tab:kepler_results}. We found that KSN 2011a is best fit by a $n=3$ model, appropriate for a BSG progenitor. We took $f_\rho=0.1$, $\kappa_{0.34}=1.$, and $T_{col}=1.0$ (see discussion in Section \ref{sec:model} on the choice of parameters). It was necessary to increase the errors by a factor of $1.85$ in order to achieve a best fit with $\chi^2/dof=1$. The best fit is shown in Figure \ref{fig:ksn11a_bestfit}. We did not find acceptable fits to $n=3/2$ models appropriate for RSGs. We find $R_s < 7.7 + 8.8 {\rm (stat)} + 1.9 {\rm (sys)}R_\odot$ and $\vs > 4.7-1.2 {\rm (stat)} - 1.4 {\rm (sys)}$.\footnote{The lower limit is due to the unknown extinction.}

\begin{deluxetable*}{ccccc}
	\tabletypesize{\scriptsize}
	\tablecaption{Fit results for KSN 2011a and KSN 2011d.\label{tab:kepler_results}}
	\tablewidth{0pt}
	\tablehead{\colhead{SN} &\colhead{Progenitor}& \colhead{$R_s / R_\odot$} & \colhead{$\vs$} &  \colhead{$t_0-2454833.0$}}
	\startdata
	KSN 2011a & BSG & $< 7.7 + 8.8 {\rm (stat)} + 1.9 {\rm (sys)}$ & $> 4.7-1.2 {\rm (stat)} - 1.4 {\rm (sys)}$ & $934.35^{+0.089}_{-0.066}$ \\
KSN 2011d & RSG & $111_{-21 {\rm (stat)}}^{+89 {\rm (stat)}} \phantom{}_{-1 {\rm (sys)}}^{+49 {\rm (sys)}}$ & $ > 1.8 - 0.3{\rm (stat)} -0.3 {\rm (sys)}$  & $1040.83_{-0.17}^{+0.09}$ \\

	\enddata
\end{deluxetable*}

We agree with G16 that a large radius RSG model does not fit the data. However, we find that a BSG model is consistent with the early-time data and does not require interaction to explain the fast rise.

Our result is in tension with the known association of Type II-P SNe with RSG progenitors in pre-explosion imaging \citep{smartt_progenitors_2009,smartt_observational_2015}. However, there is an observational bias against finding stars with small radii bacause they are fainter. Two additional factors suppress the detection of BSG progenitors: most progenitor detections are done with HST with red filters \citep{smartt_observational_2015}, and extinction preferentially suppresses blue stars. Half of the Type II SN that have high-quality HST pre-explosion data do not show a progenitor \citep{smartt_observational_2015}, moreover the upper limits that have been derived assume RSG progenitors. We conclude that BSGs have not been ruled out by pre-explosion imaging as the progenitors of many II-P SNe.

It is established that many SN II have a circumstellar material \citep{niemela_supernova_1985,phillips_light_1990,garnavich_early_1994,matheson_optical_2000,leonard_evidence_2000,quimby_sn_2007,shivvers_early_2015,gal-yam_wolf-rayet-like_2014,khazov_flash_2016,yaron_confined_2017}. Some recent works \citep{gonzalez-gaitan_rise-time_2015,gezari_galex_2015} argued that interaction with CSM may explain shorter than expected rise-times for II-P SNe. They suggest that the rise is due to breakout from the CSM as opposed to shock cooling. The shock-cooling models considered in this work assume CSM contributed negligibly to the LC. While it is plausible that the simplifying assumptions of shock-cooling models may not hold, some II-P \citep{yaron_confined_2017} do fit the shock cooling models well. Perhaps the fast-rising SNe in the literature are also associated with small radius progenitors.

The model is cut off by the rapid drop in temperature ($T=0.7$ eV at 1.7 days for the best fit). Uncertainties relating to recombination, and the internal structure of the star make it difficult to assess if a small radius progenitor can or cannot support a $\sim100$ day plateau. From Figure 1 in G16, the LC is at $M\approx-15$ at 130 days indicating a \Ni mass of $0.06$ M$_\odot$ assuming full gamma-ray trapping. Note that the tail does not appear to follow cobalt decay, and fades at a rate closer to 1 mag per 50 days, which is unusual for a II-P. 

Following \cite{nakar_importance_2016} we can estimate the energy contributions to the light curve. The total $ET$ (time weighted energy) from this SN can be roughly estimated by taking the plateau luminosity $M_{pl}=-17 \rightarrow L_{pl}\sim 1.75\times10^{42}$ erg s$^{-1}$. Using the method presented in \cite{nakar_importance_2016} we can estimate $ET$ and find that for a plateau luminosity of $1.75\times 10^{42}$ lasting for roughly 100 days we get $ET\sim 6\times 10^{55}$ erg s. The $ET$ contribution of \Ni can be readily calculated as $2\times 10^{55}$ erg s. This leaves $\sim 4\times10^{55}$ erg s to be accounted for by cooling envelope emission. Using the relations presented in \cite{shussman_type_2016} we use their equation 10 reproduced here:

\begin{equation}
	ET \approx 0.15 E_{exp}^{1/2} M_{ej}^{1/2} R_s = 2.85\times10^{55} E_{51}^{1/2} M_{15}^{1/2} R_{500}\; {\rm erg\; s}
\end{equation}

\n Where $E_{51}=E/10^{51}erg$, $M_{15}=M/15M_\odot$ and $R_{500}=R/500R_\odot$. Typical values of ET are $\sim 0.5-7\times10^{55}$ erg s \citep{nakar_importance_2016}. There is a factor of 5-10 uncertainty in the relation from \cite{shussman_type_2016}. Therefore a radius of $10R_\odot$ induces roughly a factor of $5-10$ increase in ET, which must be explained by a factor of 100 increase in $E_{exp}M_{ej}$, i.e. $E_{51} M_{15}=100$. One factor of 2 can be absorbed in the ejected mass, leaving a factor of 50 to be absorbed in $E_{51}$. \cite{rubin_type_2016} found that $E_{51}/M_{10}$ spans two orders of magnitude (0.2-20), so such a large energy is not impossible. The \cite{shussman_type_2016} relation depends on MESA progenitors \citep{paxton_modules_2011} which may not be representative of stellar profiles just before explosion. We therefore conclude that a radius of $\sim10R_\odot$ cannot be rejected based on energy budget and plateau length considerations.

\begin{figure*}[ht]
	\centering
	\includegraphics[width=0.75\textwidth]{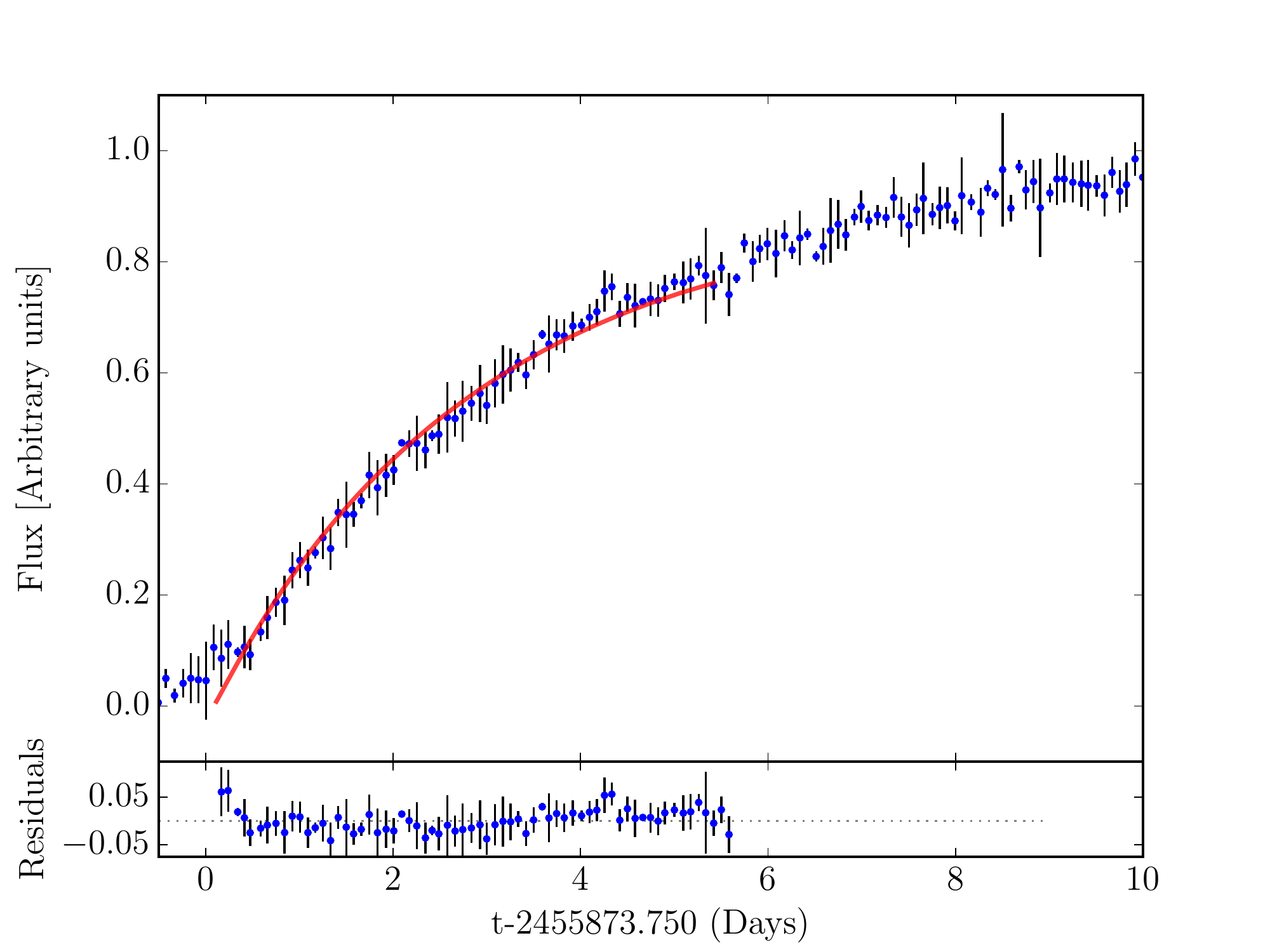}
	\caption{Top: best fit to KSN 2011d. Blue points are the 2 hour binned data. The models are valid only for the times where the best fit red line is drawn. Bottom: residuals.}
	\label{fig:ksn11d_bestfit}
\end{figure*}

KSN 2011d is best fit by an $n=3/2$ model, appropriate for a RSG progenitor. We took $f_\rho=1$, $\kappa_{0.34}=1.$, and $T_{col}=1.1$. It was necessary to increase the errors by a factor of $2.0$ in order to achieve any reasonable fits. The best fit is shown in Figure \ref{fig:ksn11d_bestfit}. We did not find acceptable fits to $n=3$ models appropriate for BSGs. We found that $R_s = 111_{-21 {\rm (stat)}}^{+89 {\rm (stat)}} \phantom{}_{-1{\rm (sys)}}^{+49 {\rm (sys)}}$ and $\vs > 1.8 - 0.3{\rm (stat)} -0.3 {\rm (sys)}$.

Our results do not agree with those of G16, however the cause is not entirely clear. We too find that an RSG model is in excellent agreement with the data, however our constraint on the radius excludes their best fit value. We are unable to recover the reported calculations of G16. The peak magnitude of their reported best fit according to RW11 equations 13-14 (despite being beyond the limit of validity) is $m_{peak}=20.0$ (including MW extinction), however using the same parameters they calculate it to be $m_{peak}=20.23$. This quarter magnitude discrepancy is the primary source of our conflicting results for the radii. 

\begin{figure}[ht]
	\includegraphics[width=1\columnwidth]{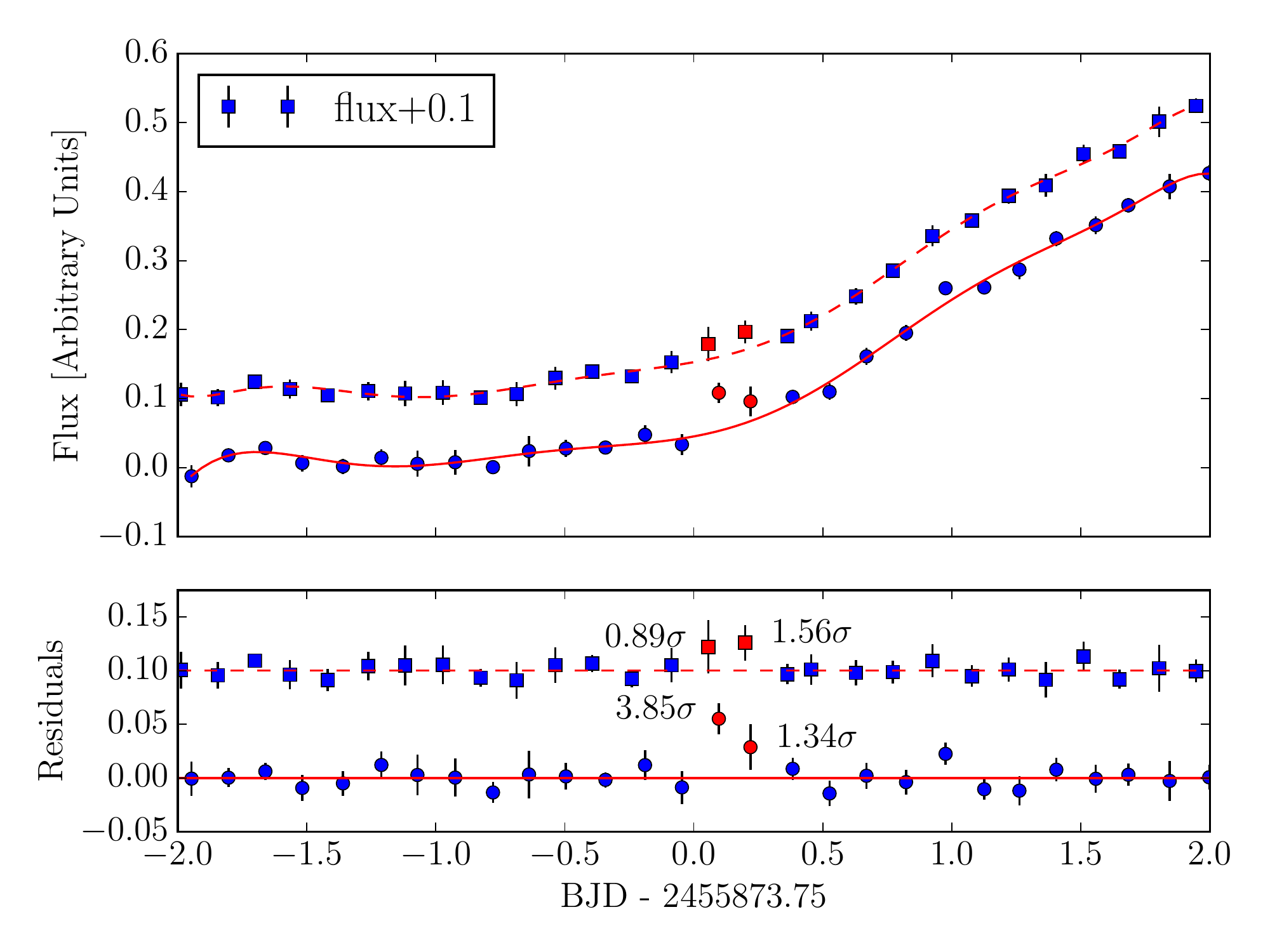}
	\caption{Significance of the departure of the shock-breakout identified by G16. Top: the early time light curve binned to 3.5 hour intervals with different binning phases, offset by $0.1$ for visual clarity. The data have been fit to a 9th order polynomial (excluding the two points in red) to test departure from a smooth function. Bottom: residuals from the smooth functions. Binning has a dramatic effect on the significance and shape of the departure, which at most is $3.85\; \sigma$, but can drop to $1.56\; \sigma$.}
	\label{fig:breakout_test}
\end{figure}

In regards to the claim of shock breakout we find that their result is not statistically significant. We examined all of the seven possible to bin 0.5 hr data points into 3.5 hr bins (the bin width used by G16), and present the most and least significant departures from a smooth rise in Figure \ref{fig:breakout_test}. The most significant binning option closely resembles the data presented in G16, however the least significant looks dissimilar to the shock-breakout model in shape, and is much weaker in significance. We find that the shape and significance strongly depend on the choice of binning, and conclude that the G16 result is not robust. 

\begin{figure}[ht]
	\includegraphics[width=1\columnwidth]{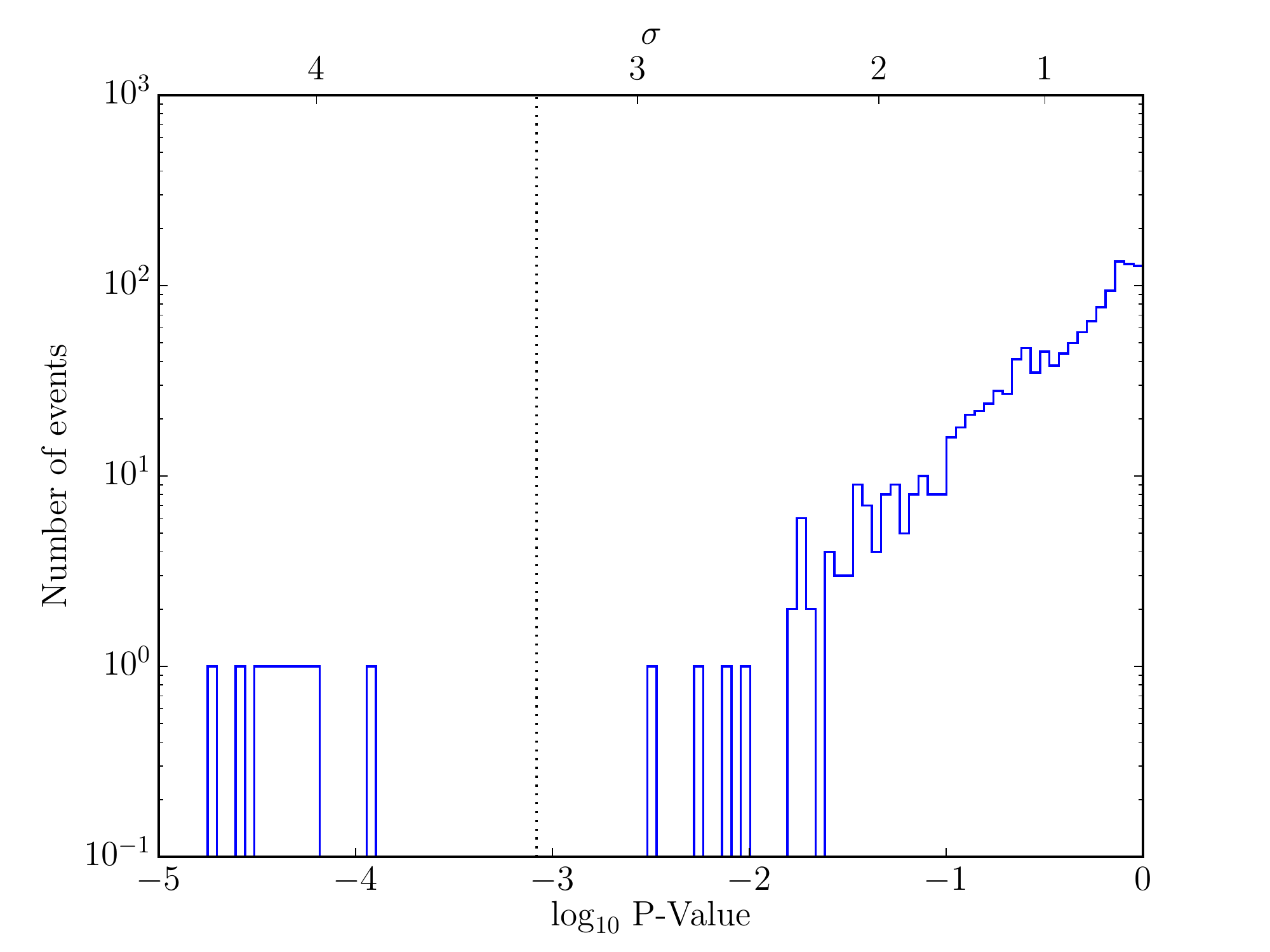}
	\caption{Histogram of P-values of ten consecutive points (not binned) with respect to the background in the data up to two days before the SN explosion. The P-value of the claimed ``shock-breakout'' is shown in the vertical dotted line. Several events have less likely departures in the noise.}
	\label{fig:prob_of_outliers}
\end{figure}

Our second test of significance shows that the Kepler data is so highly sampled that it is very probable to see $3\sigma$ and $4\sigma$ departures. Figure \ref{fig:prob_of_outliers} shows the P-value of all collections of ten consecutive points in the light curve up to two days before the explosion. Not only are these likely, but in the noise before the SN explosion there are several departures with much lower P-value (higher significance). Given the number of data points, the false alarm probability is too high to warrant a discovery claim. We conclude that G16's result is not statistically significant, and more events of this nature must be studied.

\section{Conclusions}
We have explored the uncertainties in applying SW17 shock cooling models to observations. Generating synthetic photometry with a realistic followup campaign and noise model we have shown that ultraviolet coverage is necessary to constrain the progenitors radius in a meaningful way. It is clear that ground-based campaigns will be limited in their ability to constrain the progenitors radius. Shock cooling models are discriminative with regards to the polytropic index for large radii. The uncertainties are strongly influenced by the limits of validity of the models, as was explained in \citet{rubin_type_2016}, although several works have not treated them systematically---leading to incorrect conclusions.

Multi-band light curves have the potential to constrain the local host extinction---given reasonable assumptions on $R_V$---with the best performance with high cadence ultraviolet coverage. A dedicated UV satellite such as \ul \citep{sagiv_science_2014} would provide superior coverage even to that which was explored in this work.

We applied our methods to the SN LCs recently published by G16. Our findings do not agree with theirs. First, we were unable to reproduce G16's results based on the information provided in their paper. Our estimates of the uncertainties take into account the model's limitations. We find that a $n=3$ model can be self consistently fit to KSN 2011a. This is due in part to the fact that the observed plateau begins after $T=0.7$ eV---where the model is no longer valid. 

\section{Summary}
\begin{itemize}
	\item We have presented a method for comparing shock-cooling models to photometry self-consistently taking into account the times for which the models are valid.
	\item UV coverage at early times is necessary to statistically constrain the progenitor's radius to within the systematic uncertainties.
	\item UV coverage at early times in conjunction with optical bands can constrain the local host extinction under assumption on $R_V$.
	\item The ejected mass is weakly correlated with $\vs$ and $R_s$.
	\item Both KSN 2011a and KSN 2011d can be self-consistently fit with BSG and RSG shock-cooling models respectively.
	\item The shock-breakout of KSN 2011d reported by G16 is not statistically significant and depends strongly on binning effects.
\end{itemize}

\acknowledgments
We thank E. Waxman, N. Sapir, and E.O. Ofek for helpful discussions. We thank P. Garnavich for the Kepler SN data. This research made use of Astropy, a community-developed core Python package for Astronomy \citep{astropy_collaboration_astropy:_2013} and the MATLAB package for astronomy and astrophysics \citep{ofek_matlab_2014}. We also made use of the package \emph{emcee} \citep{foreman-mackey_emcee:_2013} and corner.py \citep{foreman-mackey_corner.py:_2016}. AG-Y and AR are supported by the EU/FP7 via ERC grant No. [307260], the Quantum Universe I-Core program by the Israeli Committee for Planning and Budgeting and the ISF; by an ISF grant; by the Israeli ministry of science and the ISA; and by Kimmel and YeS awards.

\bibliographystyle{aasjournal.bst}
\bibliography{biblio}

\end{document}